# Finite-momentum nematic fluctuations soften phonons in the superconducting state of Ba(Fe$_{1-x}$Co$_x$)$_2$As$_2$

F. Weber[1], D. Parshall[2,3], L. Pintschovius[1], J.-P. Castellan[1,4], M. Merz[1], Th. Wolf[1], M. Schütt[5], R. M. Fernandes[5], and D. Reznik[3,6]

[1] Institute for Solid State Physics, Karlsruhe Institute of Technology, 76021 Karlsruhe, Germany
[2] NIST Center for Neutron Research, National Institute of Standards and Technology, Gaithersburg, Maryland 20899, USA
[3] Department of Physics, University of Colorado at Boulder, Boulder, Colorado, 80309, USA
[4] Laboratoire Léon Brillouin (CEA-CNRS), CEA-Saclay, F-91911 Gif-sur-Yvette, France
[5] School of Physics and Astronomy, University of Minnesota, Minneapolis, Minnesota, 55455, USA
[6] Center for Experiments on Quantum Materials, University of Colorado at Boulder, Boulder, Colorado, 80309, USA

**Abstract**
Nematic order is ubiquitous in liquid crystals and is characterized by a rotational symmetry breaking in an otherwise uniform liquid. Recently a similar phenomenon has been observed in some electronic phases of quantum materials related to high temperature superconductivity, particularly in the Fe-based superconductors. While several experiments have probed nematic fluctuations, they have been primarily restricted to the uniform nematic susceptibility, i.e. $q = 0$ fluctuations. Here, we investigate the behavior of finite-momentum nematic fluctuations by measuring transverse acoustic phonon modes with wavelengths of up to 25 unit cells in the prototypical Fe-based compound Ba(Fe$_{1-x}$Co$_x$)$_2$As$_2$. While the slope of the phonon dispersion gives information about the uniform nematic susceptibility, deviations from this linear behavior at finite but small wave-vectors are attributed to finite-momentum nematic fluctuations. Surprisingly, these non-zero $q$ fluctuations lead to a softening of the phonon mode below the superconducting transition temperature, in contrast to the behavior of the phonon velocity at $q = 0$, which increases below $T_c$. Our work not only establishes a sound method to probe long wavelength nematic fluctuations, but also sheds light on the unique interplay between nematicity and superconductivity in Fe-based compounds.

**Introduction**

In some quantum materials the high temperature crystal structure has a tetragonal (fourfold) symmetry, but the electrons self-organize into a state with orthorhombic (twofold) symmetry below a certain temperature [1]. Due to the analogy with nematic liquid crystals, such electronic phases are called nematic [2]. In copper oxide superconductors, they may be associated with the formation of charge stripes whose role is not yet clearly established [3-6]. In Fe-based superconductors, the nematic order can be understood as a partially melted spin density-wave, which is often observed very close to the nematic transition [7-9]. Alternatively, this symmetry lowering has also been proposed to be driven by ferro-orbital ordering [10-12]. Either way, this phase is nematic as long as there is no long-range magnetic order, which further lowers the translational symmetry of the system. The existence of such nematic phase is well supported in many families of the iron pnictides [13,14]. It also is present in a far larger part of the phase diagram in many iron chalcogenides [15].

Figure 1(a) illustrates how magnetic fluctuations associated with the spin density-wave ground state break the tetragonal symmetry in the *a-b* plane of Ba(Fe$_{1-x}$Co$_x$)$_2$As$_2$ and introduce an orthorhombic distortion via magnetoelastic coupling [16]. The coupling to anisotropic magnetic fluctuations shortens atomic bond lengths along one direction and lengthens them along the other [17-21]. It has been shown that this orthorhombicity of the atomic lattice is driven by electronic nematic order, and not the other way around[22]. Thus, the nematic order parameter has zero wave-vector ($q = 0$) and is proportional to the deviation from tetragonal symmetry [for a review see Ref. [23]].

While nematic order triggers orthorhombic order, $q = 0$ nematic fluctuations soften the shear modulus $C_{66}$,[22] which can be measured either directly by resonant ultrasound or indirectly via the Yong's modulus $Y_{110}$ in three-point bending setups [24-26]. These measurements have shown that $q = 0$ nematic fluctuations in Ba(Fe$_{1-x}$Co$_x$)$_2$As$_2$ are present for a wide temperature and doping range [Fig. 1(c)]. At chemical substitution levels for which the tetragonal-to-orthorhombic transition is suppressed, these $q = 0$ nematic fluctuations generally increase on cooling but decrease rapidly on entering the superconducting phase, signaling a competition between superconductivity and nematic order[22]. Note that the uniform nematic susceptibility has also been probed via electronic Raman scattering[27,28].

An interesting and unexplored question is about the behavior of the nematic fluctuations at finite wave-vectors $q > 0$. In this paper, we argue that the latter should couple to the transverse acoustic (TA) phonons dispersing along the [010] direction and polarized along [100]. These TA phonons correspond to the vibrational shear modes of the whole crystal in the long wavelength limit.

Our previous investigation established that in *undoped* BaFe$_2$As$_2$ and SrFe$_2$As$_2$ the TA phonons soften on approach to the nematic phase transition [29] both on heating and on cooling. The softening tracks the nematic and magnetic fluctuations in the high temperature tetragonal phase [30]. In the orthorhombic phase, these TA phonons harden proportionally to the orthorhombic order parameter (note that neither resonant ultrasound nor 3-point-bending techniques can be used to investigate the orthorhombic phase because of crystal twinning).

In our work we measured three samples of *doped* Ba(Fe$_{1-x}$Co$_x$)$_2$As$_2$ displaying qualitatively different behaviors: (1) Under-doped samples exhibiting three characteristic temperatures related to the structural ($T_s$), magnetic ($T_N$) and superconducting phase transitions ($T_c$). (2) An optimally-doped sample without structural and magnetic transition and in close proximity to a putative quantum critical point [24]. (3) An over-doped sample with strongly reduced nematic fluctuations [24,26]. At small enough momenta, the phonon dispersion is linear in $q$, and the corresponding phonon velocity is proportional to the square root of the shear modulus $C_{66}$ – and thus to the uniform nematic susceptibility. However, as $q$ increases, the dispersion becomes non-linear. As we explain below, the main contribution to this non-linearity is expected to arise from $q > 0$ nematic fluctuations, as illustrated in Fig. 1(d).

We found that nematic fluctuations intensify on cooling in the tetragonal phase but are suppressed by the nematic (structural) transition. Surprisingly, the TA phonons soften further on entering the superconducting state, in contrast to the shear modulus, which hardens below $T_c$. We show that this behavior is linked to the non-linearity of the phonon dispersion, which sets in at surprisingly small momentum transfer. This non-linearity – and thus the phonon softening – is likely a result of the effect of finite-momentum nematic fluctuations.

**Results**
Figure 2 illustrates how the softening of long-wavelength acoustic phonons appears in the raw data. We performed constant energy scans along $(2,K,0)$, $-0.1 \leq K \leq 0.1$ [scan geometry is displayed in Fig. 2(d)]. For simplicity, we refer always to wavevectors given in reciprocal lattice units (r.l.u.) of the tetragonal high-temperature structure, which is also the low-temperature structure for $x \geq 0.06$. Two peaks on opposite sides of $K = 0$ correspond to acoustic phonons propagating along $K$ and along $–K$. A tilted resolution ellipsoid and corresponding focusing effects are intrinsic to triple-axis neutron scattering measurements [31]. It can result in an asymmetric lineshape with a broad (defocused) peak on one side and a narrow (focused) one on the other, e.g. as in the optimally doped sample [Fig. 2(b)]. The degree of the asymmetry depends on the direction of the tilt of the resolution ellipsoid relative to the phonon dispersions. In our measurements of the under- and optimally doped samples [Figs. 2(a)(b)], the resolution function further picked up a small contribution from more steeply dispersing longitudinal acoustic phonons centered at $K = 0$ because of imperfect momentum resolution in the longitudinal direction. Hence, our constant energy scans were fit with three peaks with the middle peak position fixed to $K = 0$. When the phonons soften, the scan at a fixed energy crosses the dispersion curves at larger $|K|$ [see Fig. 2(d) for an illustration] and the separation between the peaks increases. This is very clear from comparison of the room temperature data and lower temperatures [Figs. 2(a)-(c)]. Effects are subtler but still visible on cooling through $T_c$.

We first discuss our results on the optimally-doped sample ($x = 0.06$), on which the most comprehensive measurements have been performed. Figure 3(a) shows that the phonon at $E = 1$ meV first softens from 300 K to about 70 K [Figs. 3(a)]. Below 70 K the temperature dependence is nearly flat down to $T_c$ where the softening resumes in the superconducting state (see also Supplementary Figure 1 and Supplementary Note 1).

The phonon softening below $T_c$ is a surprising result. Generally, for small momenta, this TA phonon dispersion $E(q)$ is expected to be linear in momentum, with the proportionality coefficient related to the shear modulus, $E(q) \propto \sqrt{C_{66}} \cdot q$. Independent measurements of $C_{66}$ in samples grown in the same lab and with similar doping levels (as determined by x-ray diffraction) reveal that although $C_{66}$ does soften upon approaching $T_c$, it hardens inside the superconducting state. This behavior is shown explicitly in Fig. 3(a) (solid blue curve) [24].

To elucidate the origin of this apparent discrepancy, in Fig. 3(b) we plot the actual phonon dispersions $E(q)$ obtained from our data. Clearly, as temperature decreases, the dispersion acquires a stronger non-linear behavior. The full dispersion can be then parametrized by the expression:

$$E(q) = \sqrt{\frac{C_{66}(T)}{\rho}q^2 + \alpha(T)q^4} \quad (1)$$

Here, $\rho$ is the material density and $\alpha$ is a phenomenological parameter that reflects the non-linearity of the phonon dispersion at non-zero $q$. We will postpone the discussion of the origin of this term to the next section, where we will argue that it most likely arises from finite-momentum nematic fluctuations.

In the fits to the phonon dispersions [lines in Fig. 3(b)] we used the $C_{66}(T)$ values reported quantitatively by resonant ultrasound near optimal doping [26]. A difference in determining the doping value $x$ between samples used in Ref. [26] and samples grown in our lab has already been discussed in Ref. [24] and we used $C_{66}$ data shown to be near optimal doping while the nominal doping value of 0.084 is different from ours, i.e. $x = 0.06$. The fits are of reasonably good quality except for 250 K, where the phonon dispersion is closer to linear. Further, we can extract the values of $\alpha$ at different temperatures, as shown in Fig. 3(c). $\alpha$ is only weakly temperature-dependent down to 70K below which it displays a peak with the maximum just above $T_c$. It is the reduction of $\alpha$ below $T_c$ that promotes the observed phonon softening as a direct consequence of the non-linearity of the phonon dispersion.

The dichotomy between the phonon softening at zero momentum and at finite momentum is illustrated in Figure 3(d). Clearly, the phonon softening below $T_c$ increases as $q$ approaches zero as indicated by the solid line. Yet in the $q = 0$ limit, resonant ultrasound [26] and 3-point-bending measurements [24] reveal a steeper dispersion, i.e. hardening. Thus, the phonon renormalization below $T_c$ has to switch sign on approaching zero momentum, as indicated by the dashed line connecting to the hardening of phonons very close to $q = 0$ visible in $C_{66}$ [square in Fig. 3(b)] [24,26]. Hence, there has to be a maximum relative softening at a certain wave vector $q_{max}$ and a vanishing effect at $q_0$. Obviously, their precise values are not determined in our study. Still we indicate them in Fig. 3(d) for discussion.

Having established the behavior of the phonon softening in the optimally doped composition, in Figures 4(a) we show the results for under-doped samples. In this case, where the structural and magnetic transition temperatures $T_s$ and $T_c$ are well separated ($\Delta T \approx 10$ K), the phonon first softens on cooling from room temperature but then hardens between $T_s$ and $T_c$, essentially following the temperature dependence of the orthorhombic order parameter [solid red line in Fig. 4(a)]. This relationship between the phonon hardening below $T_s$ and the orthorhombic order parameter was reported for undoped $BaFe_2As_2$ and $SrFe_2As_2$ [30]. On cooling below $T_c$ the temperature dependence of the phonon softening is again reversed. Qualitatively, this could be explained by a similar reversal of the orthorhombic order parameter, which was also reported in the literature [13]. However, when the temperature dependencies of the phonon softening and of the orthorhombic order parameter are placed together, the phonon softening in the superconducting state seems to be larger than the order parameter suppression [Fig. 4(a)]. Thus, superconductivity may have a similar effect on the phonon softening as in the optimally-doped sample.

In the overdoped sample ($x = 9.5$ %) the initial hardening for $E = 1.5$ meV [Fig. 4(c)] on cooling from room temperature reflects the conventional behavior when nematic fluctuations are negligible. Below $T = 150$ K, nematic fluctuations take over and the phonon starts to soften. This softening accelerates through $T_c$ with no clearly observable plateau right above $T_c$. The softening for $E = 1.5$ meV continues below $T_c$ with only a small temperature region, $T_c = 17\,K \leq T \leq 35\,K$ exhibiting a slightly reduced softening. Results for an energy transfer of $E = 2.5$ meV (Supplementary Figure 1) show a general softening on cooling below room temperature with no detectable change of the evolution across $T_c$.

**Discussion**

As presented above, the observed phonon softening near the superconducting transition temperature does not arise from a softening of the sound velocity, but rather from non-linear effects in the TA phonon dispersion. The main question is about the microscopic origin of the non-linear parameter $\alpha$, whose behavior encodes the phonon softening below $T_c$. While a full analysis certainly requires additional experimental and theoretical input, important qualitative points can be made.

Formally, the renormalization of the phonon dispersion arises from the phononic self-energy Feynman diagram shown in Fig. 1(b). While the coupling between TA phonons and electrons vanish in the simplest form of the electron-phonon Hamiltonian, Umklapp processes and disorder effects promote a non-zero coupling. As a result, in a non-interacting electron gas, this self-energy correction to the phonon dispersion is proportional to the standard Lindhard function $\Pi$. At temperatures well below the Fermi temperature, as it is presumably the case near $T_c$, the Lindhard function depends on the ratio $E(\mathbf{q})/\mathbf{q}$, implying that the phonon dispersion remains linear. Thus, it is unlikely that the pronounced temperature dependence of the non-linear coefficient $\alpha$ near $T_c$ arises from non-interacting effects.

Among the several interactions present in the system, those associated with nematic fluctuations are the most natural candidates to explain the behavior of $\alpha$. First, symmetry imposes a bi-linear coupling between nematic order and orthorhombic distortion. Second, the softening of the shear modulus $C_{66}$, which determines the slope of the TA phonon dispersion in the limit of small wave-vectors, is primarily governed by the uniform nematic susceptibility. Thus, it is natural to consider that the non-linear coefficient $\alpha$ is strongly affected by nematic correlations at finite momenta. The latter can arise from both the finite-momentum nematic susceptibility and the momentum dependence of the nematic coupling constant, which is determined by the $\mathbf{q}$-dependence of the phonon eigenvector [Fig. 1(d)] and of the coherence length of nematic fluctuations. Disentangling these two contributions require additional theoretical calculations beyond the scope of this work.

One can also ask if standard electron-phonon coupling could cause the observed phonon softening. In fact, there have been many reports of anomalous properties of TA phonons in conventional, i.e. phonon-mediated superconductors [32-35]. In fact, the temperature dependence of TA phonons in elemental niobium scales qualitatively nicely with the results obtained at 2.5 meV in the optimally doped sample [Supplementary Fig. 1(b)]. However, in niobium the respective phonons are at an energy about 10% higher than the superconducting gap $2\Delta_{Nb} =$ 3.2 meV. By all reports, any combination for $\Delta_1/\Delta_2 + \Delta_2/\Delta_1$ (allowing for different gap sizes on different parts of the Fermi surface connected by the phonon wavevector) in Ba(Fe$_{1-x}$Co$_x$)$_2$As$_2$ is expected to be larger than the phonon energies investigated here. Furthermore, density-functional-perturbation-theory (DFPT) calculations claim that electron-phonon coupling is very weak and unimportant in Fe-based superconductors with regard to the high superconducting transition temperatures [17]. Yet, DFPT predicts also the splitting in energy of certain degenerate phonon modes on going from the high-temperature tetragonal to the low temperature orthorhombic unit cell [36]. Such splitting was recently measured to be much smaller than predicted [37]. DFPT calculations work in the quasi-harmonic approximation, and their quantitative disagreement with experiment may reflect the suppression of ordered moment by fluctuations. The phonons that we investigated here are not sensitive to the ordered moment but rather to fluctuations that are not included in DFPT. So, invoking DFPT is of limited utility in explaining our results.

It is surprising that the non-linear contributions to the phonon dispersion, which cause a strong phonon softening below $T_c$, become relevant already at very small momenta, $|q| > 0.04$ r.l.u.. This could indicate a very strong softening of the sound velocity and/or a significant enhancement of finite-momentum fluctuations that is not translated into an enhancement of the uniform susceptibility. It is interesting that this behavior is more pronounced in the optimally doped sample, which is presumably close to a putative nematic quantum phase transition[38] (see also Supplementary Figure 2 and Supplementary Note 2). Whether this enhanced non-linear behavior is connected to nematic quantum criticality is an interesting topic for future investigations. Moreover, our data implies the existence of a small momentum scale $q_0$ below which the phonon softening changes to a phonon hardening [see Fig. 3(d)]. This momentum scale, and its evolution with doping, may provide important clues about the interplay between nematicity and superconductivity.

**Conclusion**

To conclude, we showed that the suppression of low-energy nematic fluctuations observed in Ba(Fe$_{1-x}$Co$_x$)$_2$As$_2$ at zero momentum is not reflected in the behavior of the structural soft phonon mode at finite small momentum with $|q| > 0.04$ r.l.u.. The temperature dependence of the soft mode in the tetragonal non-superconducting phase is qualitatively similar to that of the fluctuations at the ordering wavevector, $q = 0$. Below the superconducting transition, however, the trends switch: resonant ultrasound [26] and 3-point bending experiments [24] show an increase of $C_{66}$, indicating TA phonon hardening in the $q = 0$ limit, whereas the phonon measurements that directly probe small $|q| > 0.04$ r.l.u. show softening. We found that the suppression of the orthorhombic order parameter by superconductivity initiates the onset of softening at $T_c$ in underdoped samples, but this softening is greater than expected just from the competition between superconductivity and orthorhombicity. In the overdoped sample, which is far from the structural instability, $T_c$ has only a very small measurable effect on the phonons with the softening below $T_c$ continuing the trend that starts well above $T_c$. The largest and less trivial effect is clearly in the optimally-doped sample, which is presumably near the region of the phase diagram in which the nematic transition temperature goes to zero.

The phonon softening in the superconducting phase of Ba(Fe$_{1-x}$Co$_x$)$_2$As$_2$ arises from a rather pronounced non-linear behavior of the phonon dispersion, whose microscopic origin is likely related to finite momentum nematic fluctuations. The surprising momentum dependence of nematic correlations revealed by our experiments needs to be addressed for a deeper understanding of the interplay of nematicity and superconductivity. More broadly, our work demonstrates that measurements of the TA phonon dispersion provide a promising framework to probe finite momentum nematic fluctuations, complementing other probes that are sensitive only to the uniform nematic susceptibility.

## Materials and Methods

The compositions of self-flux grown single crystals of Ba(Fe$_{1-x}$Co$_x$)$_2$As$_2$ grown at the Institute of Solid State Physics were determined by single-crystal x-ray diffraction and energy-dispersive x-ray spectroscopy as done in previous work in our institute[39,40]. The inelastic neutron scattering experiments were performed on the 4F2 cold triple-axis and on the 1T thermal triple-axis spectrometers at the ORPHEE reactor (Laboratoire Leon Brillouin (LLB), Saclay, France) and on the BT-7 thermal triple-axis spectrometer at NIST center for neutron research (NCNR), Gaithersburg. We investigated two underdoped samples with slightly different doping levels of $x = 4.5\%$ and $x = 4.7\%$, one optimally-doped sample, $x = 6\%$, and an overdoped sample, $x = 9.5\%$. The $x = 4.5\%$ sample was measured at NCNR and the others at LLB. For simplicity, we use the tetragonal notation for wave vectors throughout the text. Wavevectors are given in reciprocal lattice units (r.l.u.) of ($2\pi/a, 2\pi/a, 2\pi/c$) where $a$ and $c$ are the lattice constants of the tetragonal unit cell.

## Acknowledgments

We thank J. Schmalian for useful discussions and suggestions. F.W. was supported by the Helmholtz Young Investigator Group under contract VH-NG-840. M.M. was supported by the Karlsruhe Nano-Micro Facility (KNMF). D.R. was supported by the DOE, Office of Basic Energy Sciences, Office of Science, under Contract No. DE-SC0006939. R.M.F. and M.S. are supported by the U.S. Department of Energy, Office of Basic Energy Sciences, under Award No. DE-SC0012336.

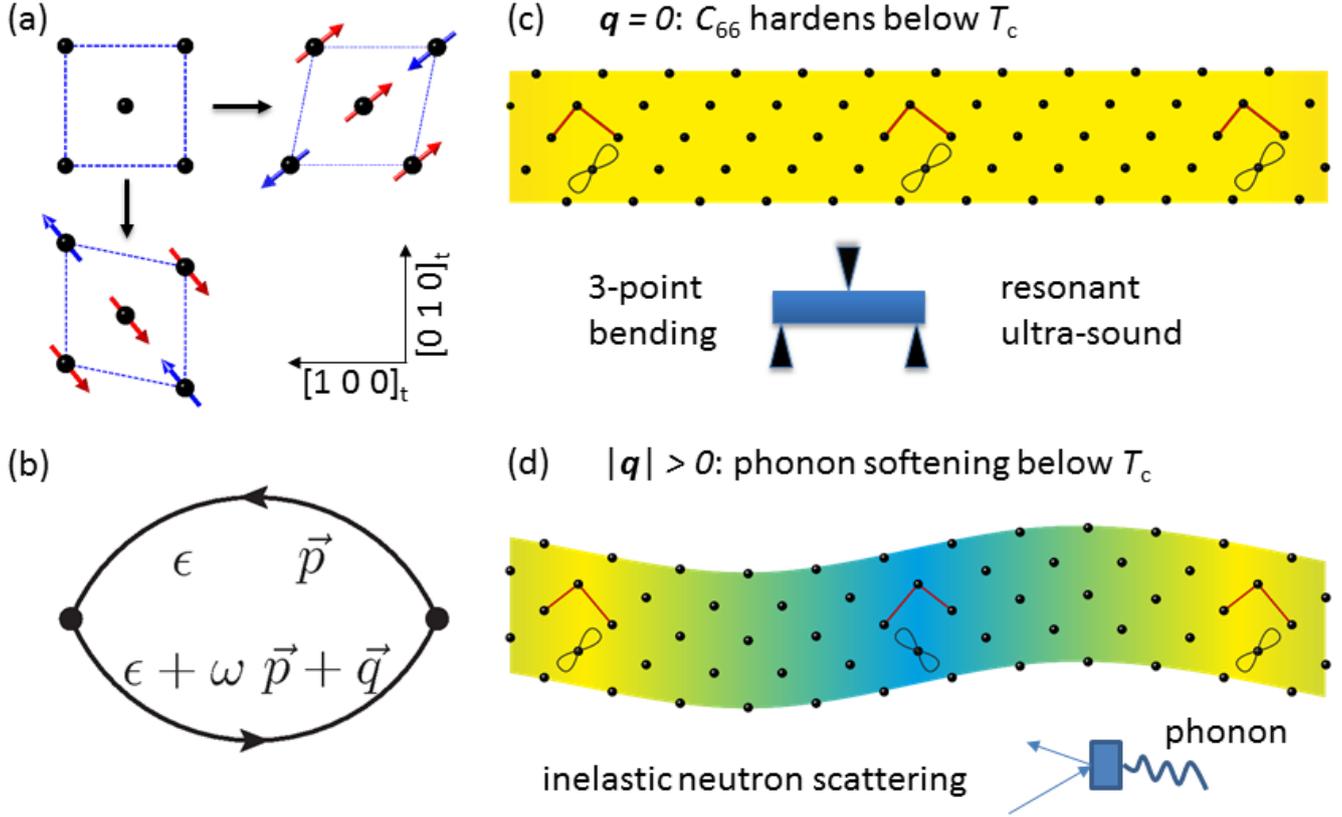

**Figure 1. Schematics of nematic fluctuations in Fe-based superconductors.** (a) Illustration of how spin fluctuations break tetragonal symmetry and induce orthorhombic distortion via magnetoelastic coupling of the atomic lattice as described in the text. (b) The phonon self-energy diagram describing how the non-interacting electron gas renormalizes the phonon dispersion. (c),(d) Nematic fluctuations in one Fe(Co)As plane of Ba(Fe$_{1-x}$Co$_x$)$_2$As$_2$. Round dots denote the Fe-atoms. Arsenic atoms are not shown. The color-coded background with blue and yellow represents opposite signs of the Ising-nematic order parameter. The red lines help visualize short and long Fe-Fe distances in the distorted lattice. The atomic arrangement in (c) corresponds to the $q = 0$ nematic fluctuations previously probed by resonant ultrasound and 3-point bending. The finite wavelength arrangement of the Fe atoms in (d) illustrates the displacement pattern of the transverse acoustic phonon (with wave vector $|q|>0$) measured by neutron scattering. While the $C_{66}$ shear modulus and, hence, the slope of the transverse acoustic phonon mode at $|q| = 0$ increases/hardens below $T_c$, we find a clear phonon softening of the same mode at $|q| > 0$.

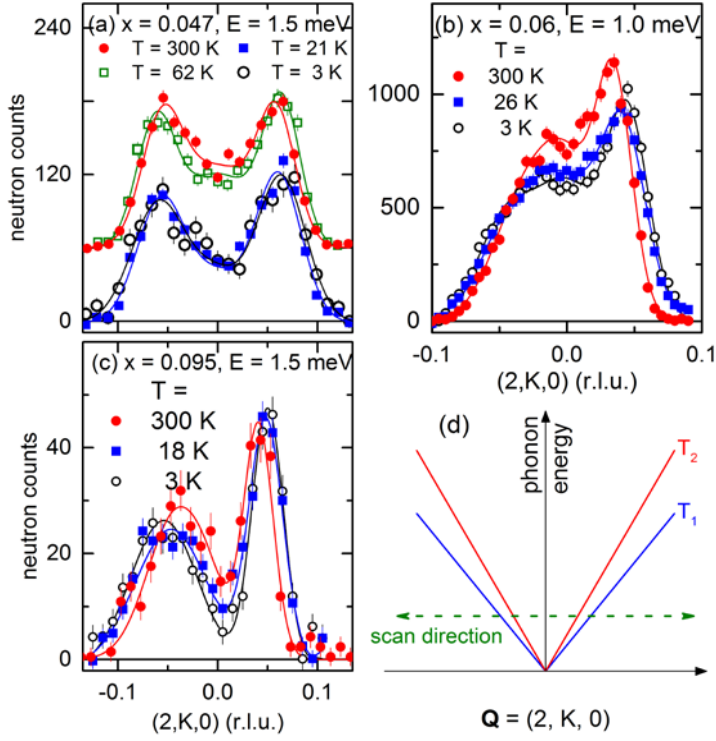

**Figure 2: Inelastic neutron scattering in Ba(Fe$_{1-x}$Co$_x$)$_2$As$_2$.** Inelastic neutron scattering data of the transverse acoustic mode in Ba(Fe$_{1-x}$Co$_x$)$_2$As$_2$ dispersing along [010] direction and polarized parallel to [100]. Measurements were done at fixed neutron energy transfer at *(a)* $E = 1.5$ meV, *(b)* $E = 1.0$ meV and *(c)* $E = 1.5$ meV for samples with $x = 4.7\%$ (underdoped), 6% (optimally doped) and 9.5% (overdoped), respectively. Estimated flat experimental backgrounds were subtracted and intensities corrected for the Bose factor. Inelastic neutron scans are shown for room temperature (dots), just above the respective superconducting transition temperatures $T_c$ (filled squares) and the base temperature (circles). Shift to higher values of *K* corresponds to a softening. Solid lines are fits to the data as described in the text. In (a) we also show data at $T = 62$ K, which is the structural transition temperature in this underdoped sample. For clarity, we added an offset of 60 neutron counts to the higher temperature data in (a). *(d)* Schematic view of the inelastic scans for the case of an acoustic phonon branch with a temperature dependent slope at temperatures $T_1$ and $T_2$. In the present case of Ba(Fe$_{1-x}$Co$_x$)$_2$As$_2$ the red line ($T = T_2$) corresponds to the dispersion at 300K in (b)(c), and the blue line ($T = T_1$) to the dispersion at low temperature.

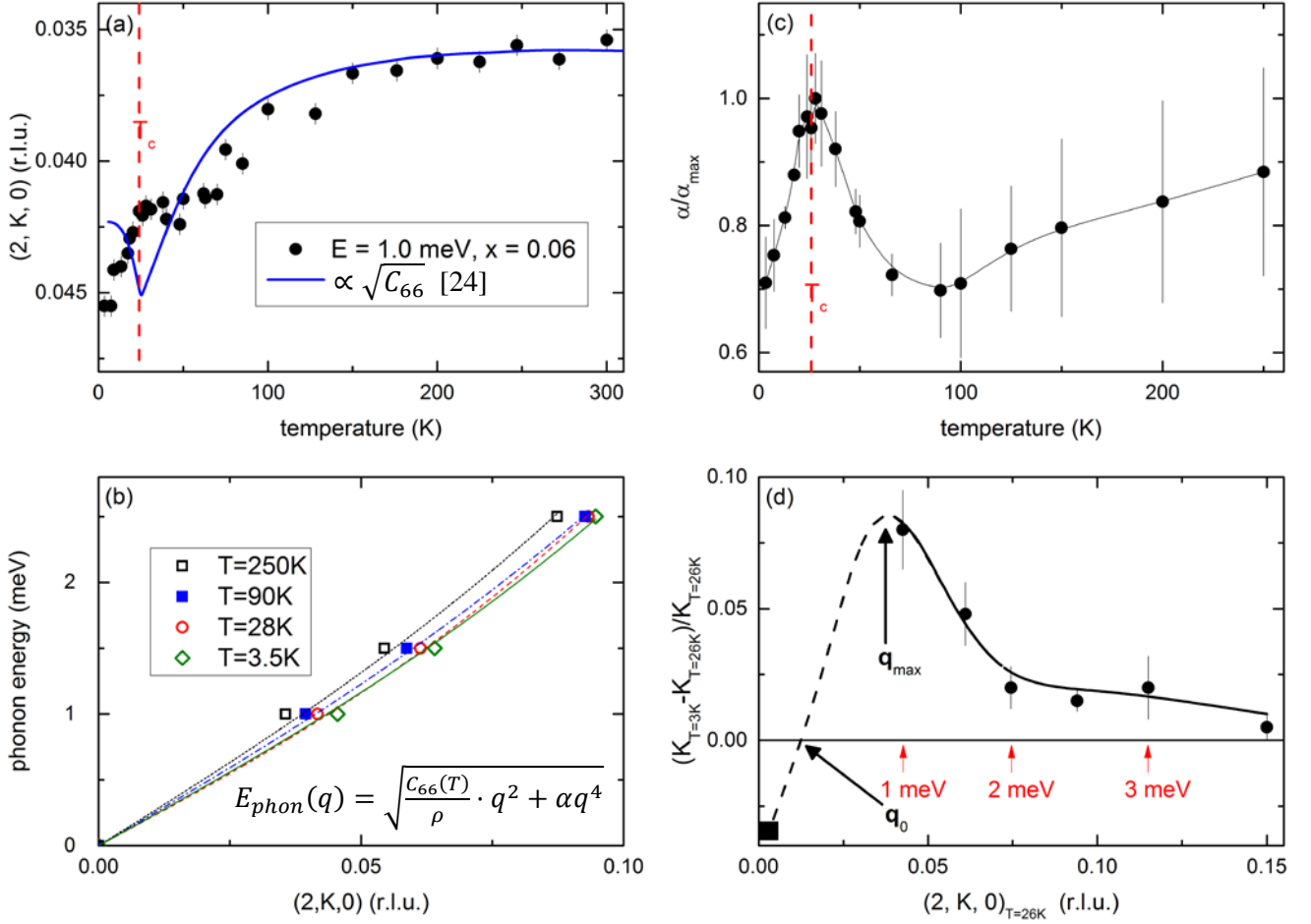

**Figure 3: Optimally doped Ba(Fe$_{0.94}$Co$_{0.06}$)$_2$As$_2$.** *(a)* Phonon renormalization in optimally doped Ba(Fe$_{1-x}$Co$_x$)$_2$As$_2$ at $E = 1.0$ meV (dots). Constant energy scans [see Fig. 2(b)] were fit to extract the phonon wavevector component $K$ in $\boldsymbol{Q} = (2, K, 0)$ at constant energies given in the panels. Larger $K$ means a less steep dispersion, i.e. softer phonons. For a more intuitive reading, we plot the $K$ scale upside-down. The solid (blue) line represents the square root of the Young's modulus $Y_{[110]}$ for $x = 6\%$, which is closely related to the $C_{66}$ shear modulus and, hence, to the slope of the TA phonon dispersion in the $K \rightarrow 0$ limit. The $Y_{[110]}$ results were taken from Ref. 24 and scaled to fit the minimum values of $K$ observed for the TA mode at $E = 1.0$ meV and $T = 300$ K. (b) Measured phonon dispersion at selected temperatures (symbols). Color-coded lines denote fits to the experimental data using the equation $E(q) = \sqrt{\frac{C_{66}(T)}{\rho} q^2 + \alpha q^4}$. For the fit we used the reported values of $C_{66}(T)$ near optimal doping (see text) [26]. $\rho$ is the sample density. (c) Temperature dependence of $\alpha/\alpha_{max}$ as deduced from the analysis shown in panel (b). (d) Relative strength of the phonon softening in the superconducting phase of optimally doped Ba(Fe$_{0.94}$Co$_{0.06}$)$_2$As$_2$ plotted versus the wavevector positions obtained at $T = 26$ K ($\approx T_c$). Red arrows with energy values indicate the constant energy of the neutron scattering scans at which the corresponding data were taken. Data in between were taken at half-integer values of energy transfer. The solid line is a guide to the eye. The dashed line is a qualitative extension to smaller $|q|$ connecting to the negative value at $K \approx 0$ (square). The latter corresponds to the increase of the $C_{66}$ shear modulus reported by resonant ultrasound [26] and 3-point bending measurements [24] for $x = 6\%$ on cooling below $T_c$ (see text). See text for the discussion of $q_{max}$ and $q_0$.

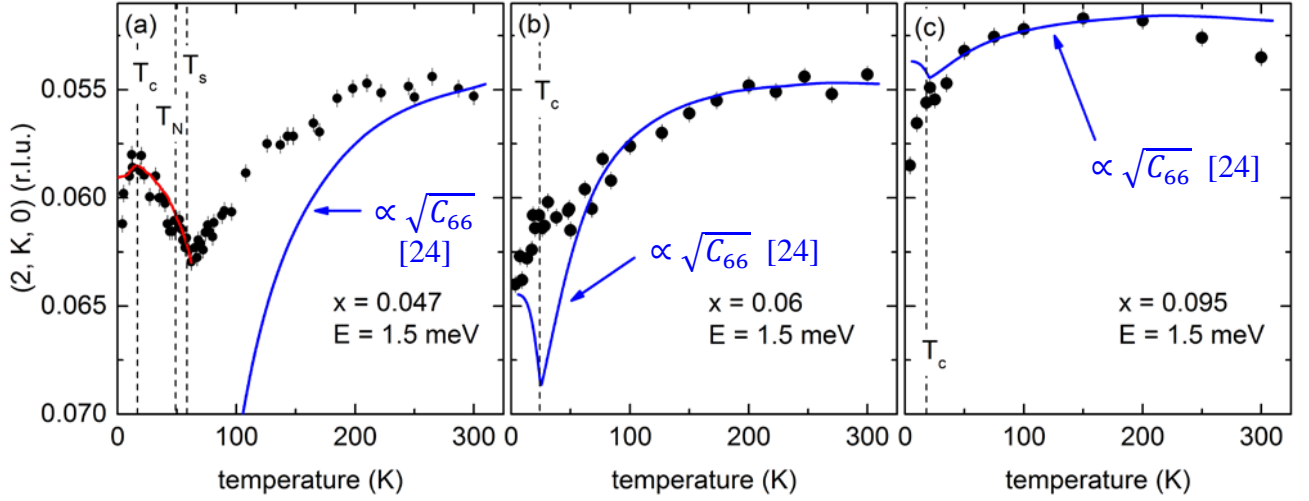

**Figure 4: Doping dependence of phonon renormalization in Ba(Fe$_{1-x}$Co$_x$)$_2$As$_2$.** Renormalization of the TA phonon in Ba(Fe$_{1-x}$Co$_x$)$_2$As$_2$ for *(a)* $x = 4.7\%$, *(b)* $x = 6\%$, *(c)* $x = 9.5\%$. Constant energy scans (see Fig. 2) were fit to extract the phonon wavevector component $K$ in $\mathbf{Q} = (2, K, 0)$ at constant energies given in the panels. Data for $x = 4.7\%$ are shown along with the respective structural ($T_s$), magnetic ($T_N$) and superconducting transition temperatures ($T_c$) indicated by the vertical dashed lines. For $x \geq 6\%$, samples have only a phase transition to a superconducting state and the respective values of $T_c$ are indicated. Larger $K$ means a less steep dispersion, i.e. softer phonons. For a more intuitive reading, we plot the $K$ scale upside-down. The red solid line in (a) is the orthorhombic splitting, which we measured by elastic neutron scattering in the sample with $x = 4.7\%$ scaled to the observed phonon hardening in the temperature region $T_c \leq T \leq T_S$. The solid (blue) lines represent the square root of the Young's modulus $Y_{[110]}$ for (a) $x = 4.3\%$, (c) $x = 6\%$ and (e) $x = 9.3\%$, which is closely related to the $C_{66}$ shear modulus and, hence, the slope of the TA phonon dispersion in the $K \to 0$ limit. The $Y_{[110]}$ results were taken from Ref. 24 and scaled to fit the minimum values of $K$ observed for the TA mode at $E = 1.5$ meV and $T = 300$ K.

**Supplementary Note 1**

**Temperature-dependence of the phonon near $Q \approx (2, 0.09, 0)$**

The data in Supplementary Figure 1 show that farther from the zone center, i.e. $q \approx (0, 0.09, 0)$, the phonon softens nearly linearly down to around 80K. The difference from the data at smaller wavevectors is due to nonlinearity of the dispersion discussed in the main text. Interestingly, the temperature-dependence at optimal doping is similar to that of a transverse acoustic phonon mode in Nb, although we argue in the main text of the paper that the similarity is probably superficial.

**Supplementary Note 2**

**Bragg scattering reflecting $q = 0$ fluctuations in optimally doped $Ba(Fe_{0.94}Co_{0.06})_2As_2$**

The intensity of the (2,0,0) Bragg reflection in $Ba(Fe_{0.94}Co_{0.06})_2As_2$ exhibits a sharp maximum located at $T_c$ [Supplementary Figure 2(a)], although a splitting as present in the under-doped samples at the structural phase transition was absent. The intensity of the (2,0,0) Bragg peak first gradually increases on cooling from room temperature as expected from the diminished thermal motion of the atoms. However below 80 K it increases sharply, which is mostly reversed below $T_c$.

Here, it is instructive to discuss the Bragg scattering intensity for samples with varying mosaic. The intensity on a Bragg peak in a neutron scattering experiment is proportional to the number of neutrons that match the Bragg condition for a particular orientation of the samples. Since the neutron beam has a finite divergence on the order of 1 degree, Bragg scattering from a perfect crystal, e.g. Si, has low intensity, since only few neutrons are incident on the sample at exactly the right angle. Single crystals of materials other than silicon fall short of structural perfection, and have a finite mosaic spread. In this case a larger number of neutrons in a finite solid angle, that is equal to the crystal mosaic can Bragg scatter. When the crystal mosaic is much smaller than the divergence of the neutron beam, the Bragg intensity is proportional to the solid angle defined by the mosaic.

On approach to a structural phase transition critical slowing down of the fluctuations increases the mosaic of the crystal on the time scale of the neutron scattering experiment and gives rise to increased Bragg scattering intensities on top of the increase expected due to reduced thermal atomic motions expressed in the Debye-Waller factor.

Quantitatively, we can isolate the precursor effect of nematic fluctuations in our optimally doped sample by taking the difference between the observed intensities and the expected ones based on measurements of the (6,0,0) Bragg reflection, which mostly reflects the evolution of the Debye-Waller factor [see solid line in Supplementary Figure 2(a)]. We assign the intensity difference [Supplementary Figure 2(b)] to nematic fluctuations and, hence, use the term of nematic Bragg intensity. Comparison with results for the shear modulus $C_{66}$ from 3-point-bending experiments [solid (blue) line in Supplementary Figure 2(b)][1] corroborate our assignment. In particular, the decrease of Bragg scattering intensity on cooling into the superconducting state strongly suggests that our elastic measurements indeed reflect the competition between orthorhombicity and superconductivity and, hence, the influence of nematic fluctuations at $q = 0$.

Analogous measurements in the overdoped sample show no clearly observable feature at or above $T_c$ [red circles in Supplementary Figure 2(a)]. These results confirm that our optimally-doped sample is close to the quantum-critical point with enhanced nematic fluctuations at low temperature, which are suppressed below $T_c$.

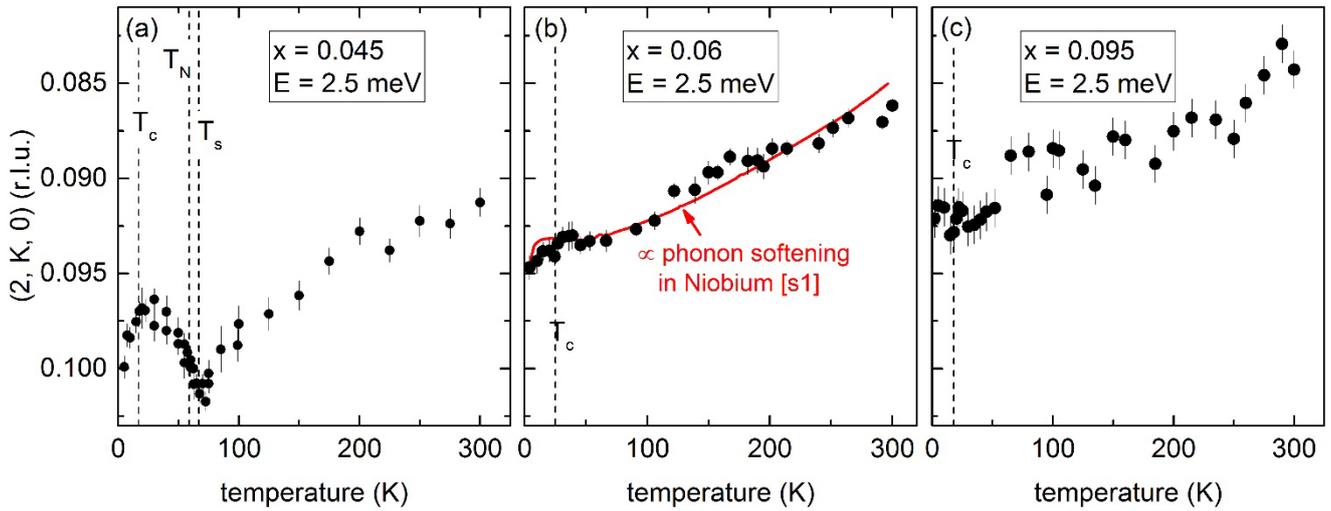

**Supplementary Figure 1: Phonon softening in Ba(Fe$_{1-x}$Co$_x$)$_2$As$_2$ observed at E = 2.5 meV (dots) for (a) x = 4.5%, (b) x = 6% and (c) x = 9.5%.** Data in (a) were taken on a different sample than those presented in Fig. 4(a) with slightly different doping values and transition temperatures indicated by the vertical dashed lines. Data shown in (a) were measured on the BT7 spectrometer at NIST center for neutron research. Data shown in (b) and (c) were taken on the same samples as those presented in Fig. 4(b) and (c) and all data sets were taken at Laboratoire Léon Brillouin, CEA Saclay. The solid red line in (b) represents qualitatively the softening of a TA phonon mode in elemental niobium[s2] scaled to fit the absolute size of the softening observed in optimally doped Ba(Fe$_{0.94}$Co$_{0.06}$)$_2$As$_2$.

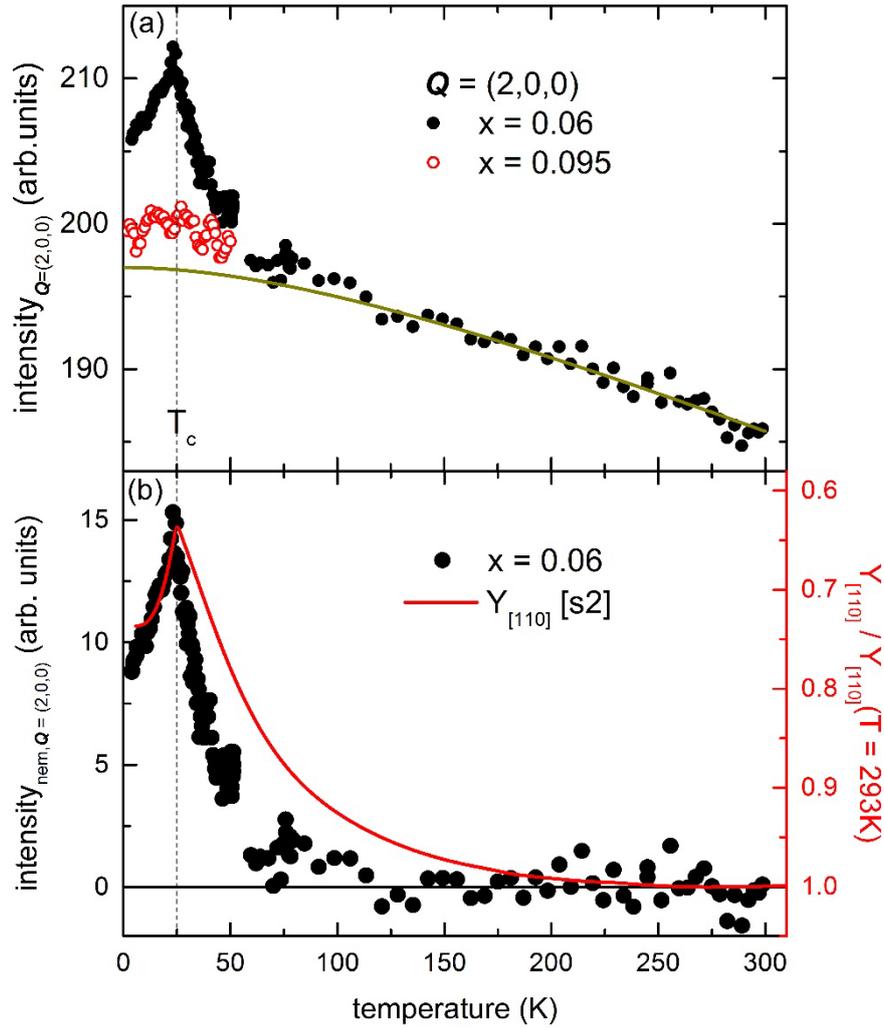

**Supplementary Figure 2: Bragg scattering reflecting $q = 0$ fluctuations in optimally doped $Ba(Fe_{0.94}Co_{0.06})_2As_2$.** *(a)* Bragg peak intensity of the (2,0,0) reflection as a function of temperature for two different doping levels, i.e. $x = 6\%$ (dots) and 9.5% (circles). The solid line represents the estimated temperature dependence based on the Debye-Waller factor extracted from measurements of the (6,0,0) Bragg reflection (not shown). *(b)* Nematic intensity (*intensity*$_{nem}$) (dots) defined by the difference between the observed intensity of the (2,0,0) Bragg reflection [dots in (a)] and the expected behavior due to Debye-Waller factor [solid line in (a)]. The solid (red) line denotes the shear modulus $C_{66}$ as extracted from measurements of the Young's modulus (right hand scale).[s1]